\documentstyle[11pt,epsfig]{article}
\begin{document}
\title{Measurement of Colliding Beam Parameters with Wide Angle Beamstrahlung}
\author{G. Bonvicini, D. Cinabro and E. Luckwald \\
{\it Wayne State University, Detroit MI 48201}}
\date{\today}
\maketitle

\section{Introduction}

Machine issues at particle factories are dominated by luminosity
optimization, which is the overlap of the density functions $\rho$ 
of the two beams
over space and time. For a single beam crossing,
\begin{equation}
L = c \int dV d\tau \rho_1({\bf r},\tau)\rho_2({\bf r},\tau),
\label{eq:lumgen}
\end{equation}
where $dV$ is the volume element 
and $\tau$ is a time variable of order of the beam-crossing duration.
Optimal luminosity is achieved by perfect transverse overlap of two
equal and Gaussian beams squeezed to the limit allowed by the tune shift. 
For a single beam crossing, that reads
\begin{eqnarray}
L_0(t) & = & {N_1(t)N_2(t)c\over (2\pi)^3\sigma_x(t)\sigma_y(t)\sigma_z(t)}
\int dV d\tau e^{-(x^2/\sigma_x^2(t)+y^2/\sigma_y^2(t)+(z^2+(c\tau)^2)
/\sigma_z^2(t))} \nonumber\\
       & = &{N_1(t)N_2(t)\over 4\pi\sigma_x(t)\sigma_y(t)},
\label{eq:lumint}
\end{eqnarray}
where the $N_{1,2}$ and $\sigma_{x,y,z}$ 
are the beam populations and spatial dimensions
at any given time $t$ (which is a run-time variable of order one hour).

This formula becomes rather complex over time.
Particles are deflected 
by the other beam at each crossing, significantly affecting
the Twiss matrix of
the machine. The beam
currents $N_i(t)$ decrease due to beam lifetime also causing the machine's
Twiss matrix to drift. 
If the machine is perfectly symmetric, the transverse dimensions
will change but the beams will maintain perfect overlap.

Even symmetric machines have some degree of asymmetry,
and beams start moving independently
in the transverse plane as soon as the run starts. 
At $B$-Factories such as CESR,
PEP-II, and KEK, beams have horizontal dimensions $\sigma_y$ of order several
microns, with aspect ratios $\epsilon=\sigma_y/\sigma_x \sim 0.02-0.04.$ 
A drift of 5 microns is enough to spoil the luminosity.

A better description of the luminosity over time should be
\begin{equation}
L(t)=L_0(t)(1-w(t)).
\label{eq:waste1}
\end{equation}
$w(t)$ is the positive-defined waste parameter
due to non-instantaneous optimal overlap.  If $w(t)$ is known,
the wasted integrated luminosity is defined as
\begin{equation}
 L_w = f\int dt L_0(t)w(t),
\label{eq:wastelum}
\end{equation}
where $f$ is the machine frequency. The waste parameter can be readily 
derived from the convolution integrals,
Equations~\ref{eq:lumgen} and~\ref{eq:lumint}.
Dropping the time dependence, one gets 
\begin{equation}
w=1-{L\over L_0}.
\label{eq:waste2}
\end{equation}
The waste parameter is clearly of great interest, and one of the 
most important 
issues not only at $B$ factories but also at future linear
colliders. As soon as $w$ is non-zero, a correction should be applied
to restore optimal luminosity.  The ability to measure and reduce $w$
would make for a substantial
increase in the delivered luminosity of any machine.

Although $w$ can be defined mathematically from Equation~\ref{eq:waste1},
the beam-beam topology cannot be 
measured directly. Techniques have been developed that measure the 
transverse displacement of the centers of gravity of the 
beams by Bambade\cite{bambade} and Sagan, Sikora and Henderson\cite{sagan}.
Both of these techniques actively displace one of the beams, and monitor
the other beam to observe the strength of the beam-beam interaction.
In practice these techniques are sensitive to the relative
displacement of the beams centers from optimal beam-beam overlap. 

\begin{figure}[h]
\begin{center}
\thicklines
\epsfig{file=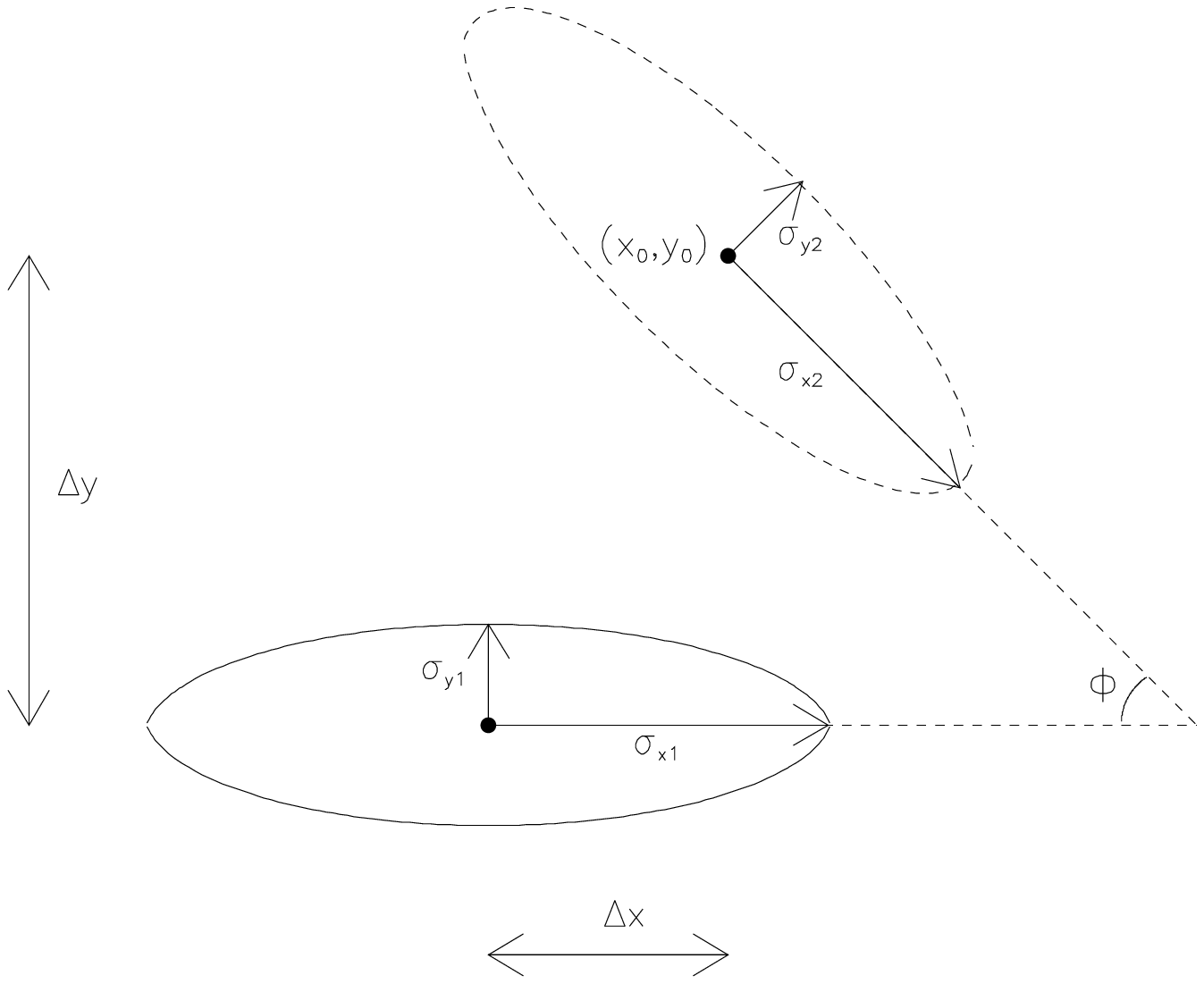,height=5.0in}
\caption{ A general beam-beam collision. Seven parameters can
be seen, corresponding to two transverse dimensions for each beam, a two
dimensional impact parameter vector connecting the two beam centers,
and one relative rotation in the transverse plane.}
\end{center}
\label{fig:dof}
\end{figure}

Generally, 
a discussion of the waste parameter must include all possible degrees of 
freedom in the evolution of a machine over a run. 
There are seven parameters
which can affect optimal beam-beam overlap\cite{welch}.
These are shown in Figure~\ref{fig:dof}.
Briefly there is a transverse displacement between the two beam
centers described by a vector $(\Delta x,\Delta y)$,
the transverse sizes of the two beams $\sigma_{x1}$, $\sigma_{y1}$,
$\sigma_{x2}$, and $\sigma_{y2}$, and a relative rotation of the two beams
$\phi$.
The two beam currents also affect the beam-beam 
interaction.  Currents are easily monitored and are
not included in the discussion below.

In this paper, a technique is proposed by which six of the seven parameters
can be passively monitored with the observation of wide angle beamstrahlung.
In case of non-zero waste, which is called a ``pathology'', the responsible
parameter is identified unambiguously,
and the amount of needed correction is measured.  The seventh
parameter can easily be measured in a beam scan also using
the wide angle beamstrahlung signal.

Seven parameters to characterize the beam-beam collision is a large number.
It is easiest to discuss the
problem if it is broken into two parts.
\begin{itemize}
\item The machine is perfectly symmetric, that is, the machine optics is
exactly the same for both beams. In Figure~\ref{fig:dof},
that means that the two
beams have zero offsets, zero rotation, and the same transverse dimensions,
resulting in only two parameters.  Dropping indices, they are the
transverse dimensions $\sigma_x$ and $\sigma_y$. If the machine is symmetric,
beams maintain optimal overlap, but the optics is affected by the
varying currents.  The luminosity is determined by
the transverse size of the beam.  The case of measuring the transverse beam
size is discussed in Section~3.
\item The beams move independently
in the transverse plane due to machine asymmetry decreasing overlap 
and luminosity.  In section~4 the measurements of the
relative sizes of the two beams, 
their transverse displacement, and
the angle between them are described. 
\end{itemize}
In this paper the use of large angle beamstrahlung,
which is described in detail in reference\cite{welch}, is described
as a beam-beam monitor that allows complete control
over both the beam-beam interaction strength and transverse displacements.
Large-angle beamstrahlung
observables, combined in a simple 2-dimensional
diagram which is called the beamstrahlung diagram, monitors
the wasted luminosity.

In Section 2 the information content of large angle beamstrahlung is 
discussed. Section 3 covers
the symmetric machine case, concentrating
on measurements of the beam size. Section 4 covers asymmetric machines,
introduces the beamstrahlung diagram, and shows how the waste parameter
can be measured.
The use of the beamstrahlung diagram to eliminate wasted luminosity
is shown in Section 5.
Three appendices are included for completeness.
Appendix A derives in a simple way three crucial properties of large
angle synchrotron radiation.
Appendix B provides a description of the beam-beam simulation developed for
this paper and Appendix C evaluates the simulation's accuracy.

\section{Large Angle Beamstrahlung}

The properties of large angle radiation, emitted
by a ultra-relativistic particle,  differ dramatically
from the classical synchrotron radiation formulae\cite{jackson}.  
Appendix A shows
that the approximations used in reference \cite{welch} and in this paper
are valid at large angles for all present and proposed 
$e^+e^-$ colliders, if beamstrahlung detection is
to be done at or near the ``magic angle'' described in \cite{welch}.
Three properties of large angle radiation are derived in Appendix A.
Of particular interest is the 100\% linear polarization,
either parallel or perpendicular to the bending force, 
obtained at certain azimuthal locations at large angle. 

At CESR for example,
it is possible to detect such radiation in visible light at a
location 5 meters away from the interaction point, at a 6mrad angle.
The beam-beam interaction occurs over a volume of order 
$300\mu$m$\times 7\mu$m$\times 7$mm, and particles are typically
deflected laterally by $10^{-2}$ mrad. Thus the light detector 
is seen at the same angle by all of the beam, and throughout the dynamic
beam-beam collision.  These are the conditions termed as ``CESR
conditions'' and used for the 
calculations of Sections 4 and 5.
A fixed fraction of the beamstrahlung energy is collected
at such a location, effectively measuring the total energy up to a constant.
Different polarization components can also be easily observed, by
filtering the observed light through polarimeters. 
 
The two polarization components can be used to build the 
radiation vectors ${\bf U}_1$ from one beam and ${\bf U}_2$ from the other
beam, which
are two-dimensional vectors in the first
quadrant.  The first dimension is the horizontal component of
the polarized beamstrahlung power signal and the second is
the vertical. The total energy vector ${\bf U}$ is defined as 
${\bf U_1+U_2}$. 
At large angles the polarization components 
and radiation spectrum
factorize\cite{albert} and a different orientation of the polarimeters
would simply rotate the horizontal and vertical axes.. 

As mentioned in the Introduction,
at present and proposed machines, beams are very flat 
($\epsilon\sim 0.02-0.04$). It is convenient to develop the
theory only for flat beams which leads to two simplifications.
First, terms of order $\epsilon$ and higher can be neglected in equations
as needed.  Second, a natural preferred orientation exists in the
transverse plane, which is adopted to produce the results of this paper. 

It should be noted that
two counters on each side, each looking
at a different polarization component, and in absence of background, 
are enough to extract complete information
from beamstrahlung.  As an example, given the formulae in Appendix A,
$U_x$ can be measured by measuring the $x-$polarized component at zero
degrees in azimuth, and $U_y$ by the $x-$polarized component at 45 degrees.

\section{Symmetric Machines}

If a machine is perfectly symmetric, the beam currents and transverse
dimensions of the beams will evolve, while maintaining perfect overlap.
Measurements of the beam sizes $\sigma_x$ and $\sigma_y$, determine
the luminosity.  In this case most of the interplay between machine and
beam-beam interaction is through the dynamic beta effect.

The dynamic beta effect is proportional to the average electric field
seen by one particle over many beam crossings, hence it is
proportional to the charge in the other beam, times the average 
inverse impact parameter $b$ between particles of beam~1 and particles
of beam~2\cite{jackson1},
\begin{equation}
 <{\bf E}_1>\propto N_2 <{{\bf b}\over  b^2}>.
\label{eq:E1}
\end{equation}
From\cite{albert} the beamstrahlung energy
is proportional to
\begin{equation}
U_1\propto N_1 <E_1^2>,
\label{eq:U1}
\end{equation}
or
\begin{equation}
 <E_1^2>\propto U_1/N_1.
\label{eq:EU1}
\end{equation}
The $<E_1>$ and $<E_1^2>$ are related through the transverse shape
of the beam, which can be taken to be Gaussian with no loss of
precision. Therefore,
monitoring the dynamic beta effect can be done efficiently by
monitoring the $U_i$ and the $N_i$ at the same time.

Equation~\ref{eq:U1} can be rewritten as\cite{welch,albert}
\begin{equation} 
U_1\propto {N_1N_2^2\over \sigma_x^2\sigma_z}f(\epsilon).
\label{eq:U2}
\end{equation}
The beam length, $\sigma_z$, is usually constant, and will not be considered 
here, but
clearly a beamstrahlung detector can also be used to monitor the beam length,
for example during machine studies.
The function $f(\epsilon)$ varies slowly
\begin{equation}
f(\epsilon)\sim 1+11.4\epsilon,
\label{eq:eps}
\end{equation}
and can be considered nearly constant in the following.

The result above assumes ``stiff'' beams.
A stiff beam is one where the beam particles do not change
their transverse position
appreciably during the collision.  Appendix B shows that dynamic
effects are negligible.

In flat beams
most of the impact parameter is due to the distance in $x$
between the particles, and
the energy radiated is almost only dependent on $\sigma_x$.
For perfect overlap of stiff Gaussian beams the energy $U$
is unpolarized\cite{albert}. No information
can be extracted out of polarization,
and beamstrahlung cannot monitor passively symmetric changes in $\sigma_y$.
The total power radiated is thus sensitive to $\sigma_x$.

\begin{figure}[h]
\begin{center}\thicklines
\epsfig{file=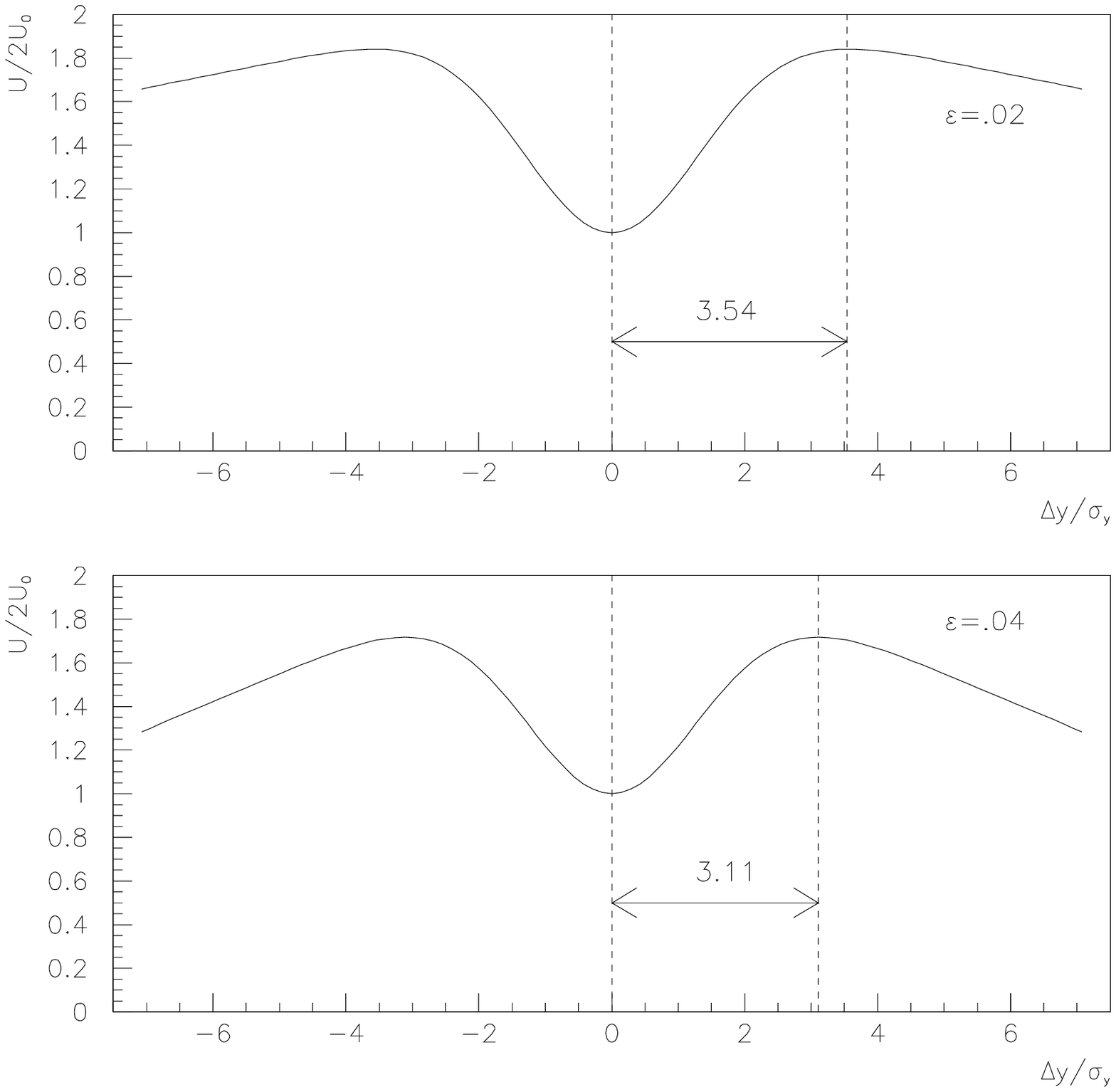,height=5.0in}
\vspace{0.25in}
\caption{ Normalized power emitted in beamstrahlung, as a function
of normalized $y-$offset. a) $\epsilon=0.02$. b) $\epsilon=0.04$.
The distance from minimum to maximum is shown, in units of $\sigma_y$.
$U_0$ is defined in Section 4.}
\end{center}
\label{fig:powvsyoff}
\end{figure}

However, as pointed out in references\cite{welch,albert}, a scan of one beam
along the vertical axis will produce the characteristic camelback feature
in the plot of $U$ versus the beam-beam offset seen in
Figure~\ref{fig:powvsyoff},
which has already been used in the detection of beamstrahlung\cite{field}.
The $\sigma_y$ can be precisely determined by 
measuring the peak-valley distance $d$ shown in Figure~\ref{fig:powvsyoff}.
The relation between $d$ and $\sigma_y$ is
\begin{equation}
 d\sim 3.97\sigma_y(1-5.4\epsilon).
\label{eq:dvssigy}
\end{equation}
Currently, the CESR beams are artificially
perturbed with an amplitude of order $0.01\sigma_y$
to measure the beam-beam interaction by observing the
effect of the perturbation on the other beam via the
lock-in effect\cite{sagan}.  It is conceivable that this
technique could ultimately
be used to determine $\sigma_y$ without scanning.  Note that a beam
scan could also be used to measure $\sigma_x$ separating it from 
$\sigma_z$.

A beamstrahlung monitor can be very useful even when a
machine is perfectly symmetric, allowing purely passive monitoring of the
beam-beam interaction and thus the beam length or $\sigma_x$.
It can be used to measure $\sigma_y$ in a beam scan.
The next Section, which deals with purely
asymmetric pathologies, shows that this method is truly valuable when beams
are not colliding head on in the transverse plane and may have different
transverse sizes.

\section{Asymmetric Machines}

If a machine is asymmetric, as all real machines are to some
degree, the two beams will drift independently in the 7-dimensional space
that induces luminosity waste.
For the purpose of studying asymmetric machines, a
single pass beam-beam simulation program was written.  
The program generates complex beam-beam configurations
involving all the pathologies shown in Figure~\ref{fig:dof}.
These configurations are, in principle, computable analytically
in the limit of stiff beams.
It was important also to cross-check the effects of
beam-beam dynamics, as the particles of one
beam are deflected towards the center of the other beam. The latter is
an effect that must be computed by simulation.
 
The simulation program is described in
Appendix~B. Its precision is evaluated in Appendix~C and is found to be
between 0.1\% and 0.2\%, for beamstrahlung yields, and better than
1\%, for the luminosity enhancement due to beam-beam pinching.
The nominal conditions to produce results in this paper were chosen as in
Table~\ref{tab:para} and are appropriate for $B$-Factories.
\begin{table}
\begin{center}\thicklines
\begin{tabular}{|c|c|c|}\hline
Paramter            & Symbol        & Value \\ \hline
Beam Width          & $\sigma_{x}$  & 300 $\mu$m \\
Beam Height         & $\sigma_{y}$  & 7 $\mu$m \\
Beam Length         & $\sigma_{z}$  & 1.8 cm \\
Bunch Charge        & $N$           & $11\times10^{10}$ \\
Relativistic Factor & $\gamma$      & $10^4$\\ \hline
\end{tabular}
\caption{Beam parameters chosen for the simulation results
presented here.}
\end{center}
\label{tab:para}
\end{table}
The measurement of the two beam sizes was discussed in the previous
section. The remaining five parameters are 
discussed here: two relative transverse dimensions; two offsets;
and one angle.  Present day beam position monitors have
spatial resolutions of order 20 $\mu$m, which is substantially less than
the $\sigma_x$ of these beams, and should always provide adequate overlap
along the $x-$axis leaving four pathologies of concern.  
An offset in $x$ will generate a unique 
configuration of the beamstrahlung diagram, which mirrors the one obtained
for an offset in $y$ which is discussed below,
and can be analyzed in a completely equivalent way.

\begin{figure}[h]
\begin{center}\thicklines
\epsfig{file=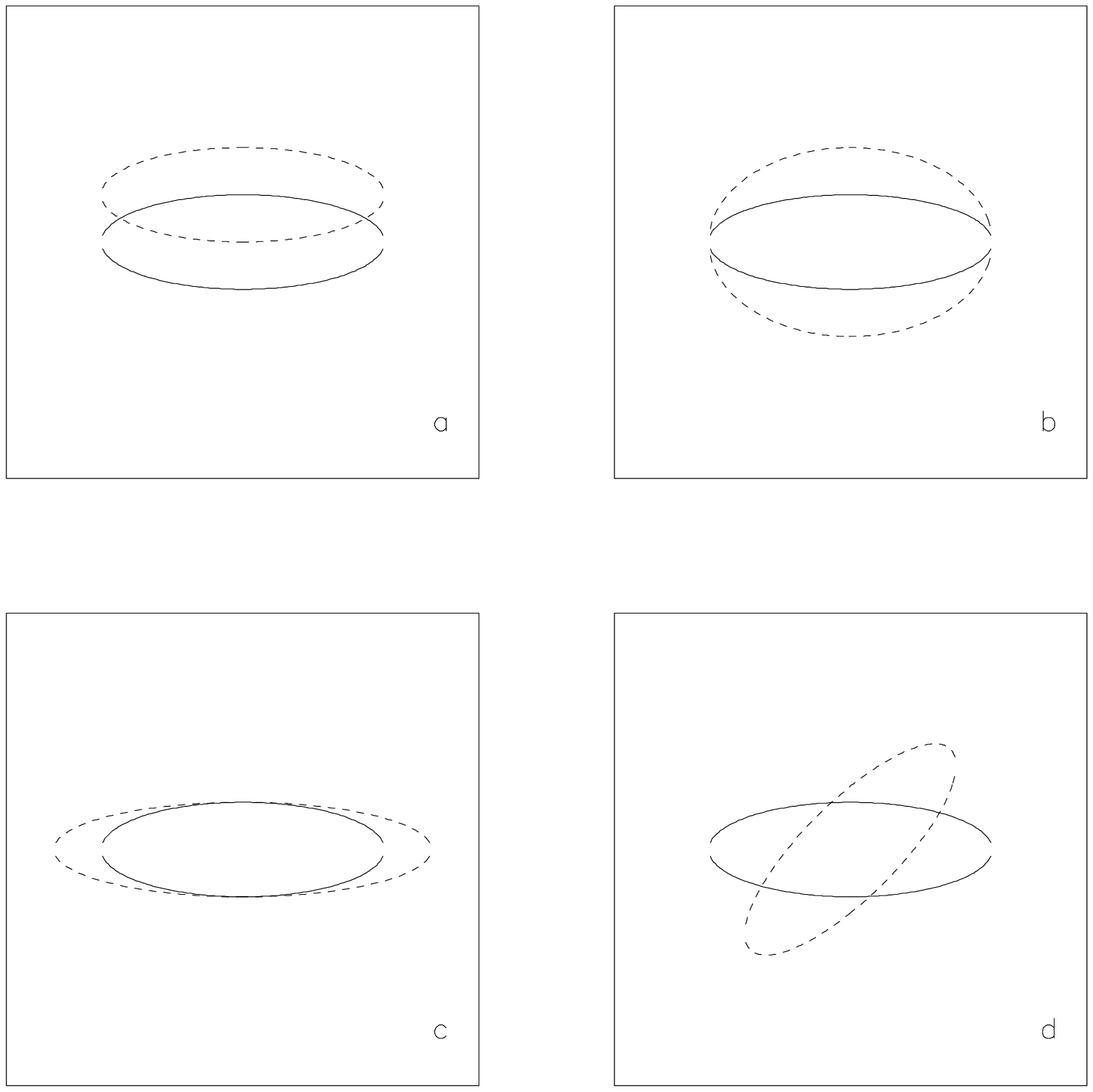,height=5.0in}
\caption{ The four beam-beam pathologies that lead to wasted
luminosity;
a) a $y-$ offset; b) $y-$ bloating; c) $x-$ bloating;
and d) a beam-beam rotation.
The pathological beam is represented by the dashed ellipse.}
\end{center}
\label{fig:4path}
\end{figure}

For simplicity it is assumed that only one beam is developing
a pathology at any given time.
The four pathologies that lead to wasted luminosity
are shown in Figure~\ref{fig:4path}.
They correspond respectively to a vertical offset, 
imperfect vertical focusing, imperfect horizontal focusing, and a
rotation of one beam with respect to the other.
All these pathologies except the third have been observed at CESR.

	The expected value of each polarization component is also assumed,
for optimal beam-beam collision, which we call $U_0$.
In practice $U_0$ can be measured experimentally by continuous observation
of colliding beams, or by separately determining the beam currents,
and $\sigma_x$ and $\sigma_y$ with a beam-beam scan as discussed in
the previous section. 

The beamstrahlung diagram 
plots ${\bf U_1,U_2}$ normalized by $U_0$.
In the figures below the contribution from
the pathological beam is represented by the dashed arrow. 
The diagram has four degrees of freedom.  The total power
monitors the beam-beam interaction strength, and
three independent dimensionless asymmetries can be defined.

\begin{figure}[h]
\begin{center}\thicklines
\epsfig{file=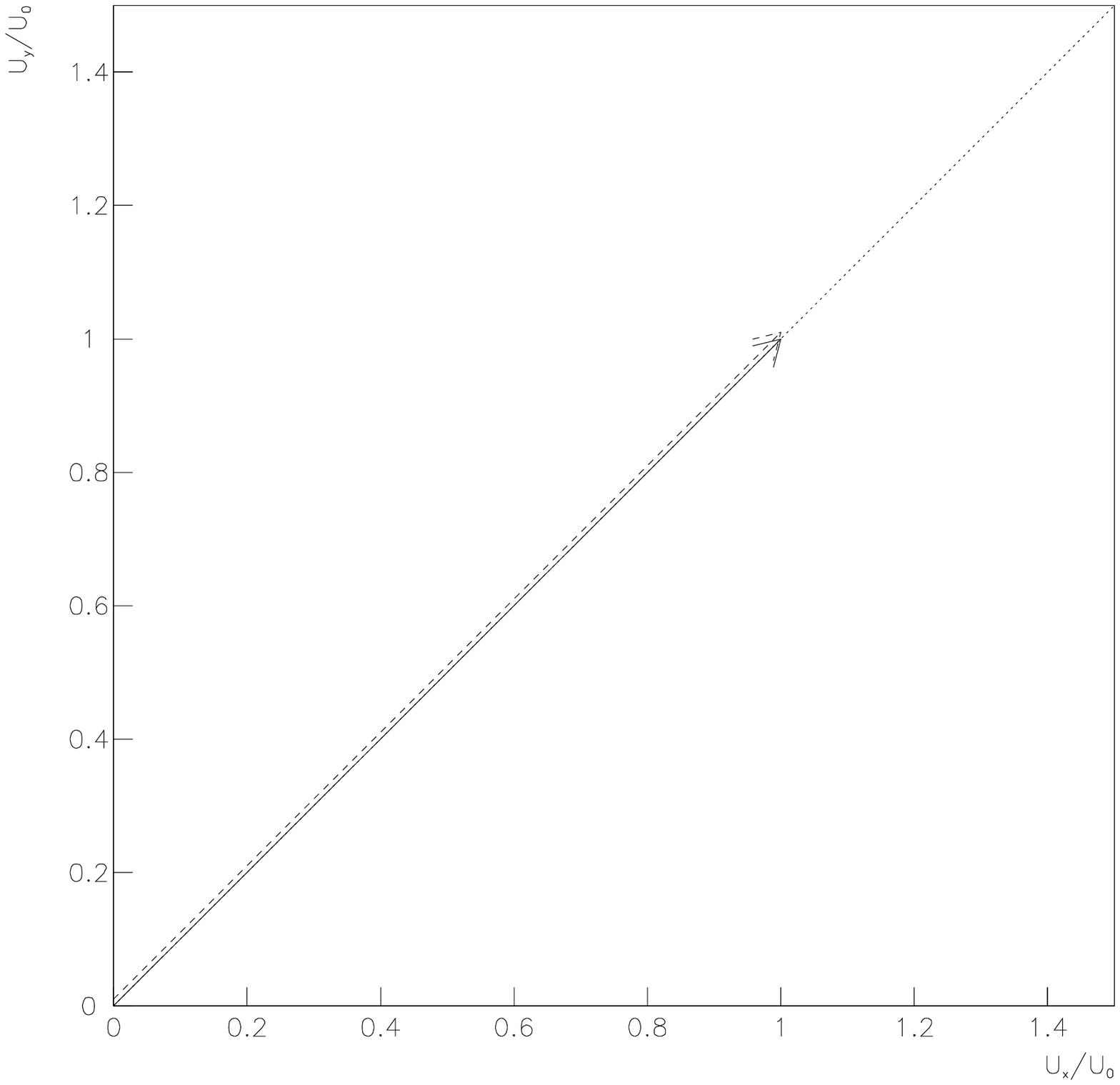,height=5.0in}
\vspace{0.25in}
\caption{ The beamstrahlung diagram corresponding to a perfect beam-beam
collision. The two vectors are exactly equal.
The dashed arrow is slightly displaced for display purposes.}
\end{center}
\label{fig:bsperf}
\end{figure}

As mentioned in Section~3 if the collision is perfect and the beams are
stiff the beamstrahlung radiation is unpolarized.  Thus
the normalized $U_i$'s are equal and the vector from each beam in
a perfect head-on collision are on top each other at 45 degrees as shown
in Figure~\ref{fig:bsperf}.
With the $U_0$ normalization one obtains the perfect collision point
at $(1,1)$ for both beams.

The effect of dynamic beams can be estimated from Table~\ref{tab:comp}
in Appendix~C.  For example at CESR dynamic beams increase $U_x$ by 0.9\% and
$U_y$ by 2.7\%, moving the perfect collision axis 
0.5 degrees above 45 degrees.  Such a small modification is nearly invisible
in Figure~\ref{fig:bsperf} and can be neglected.

Figure~5 shows for stiff beams the beamstrahlung diagrams
for each
\begin{figure}[h]
\begin{center}\thicklines
\epsfig{file=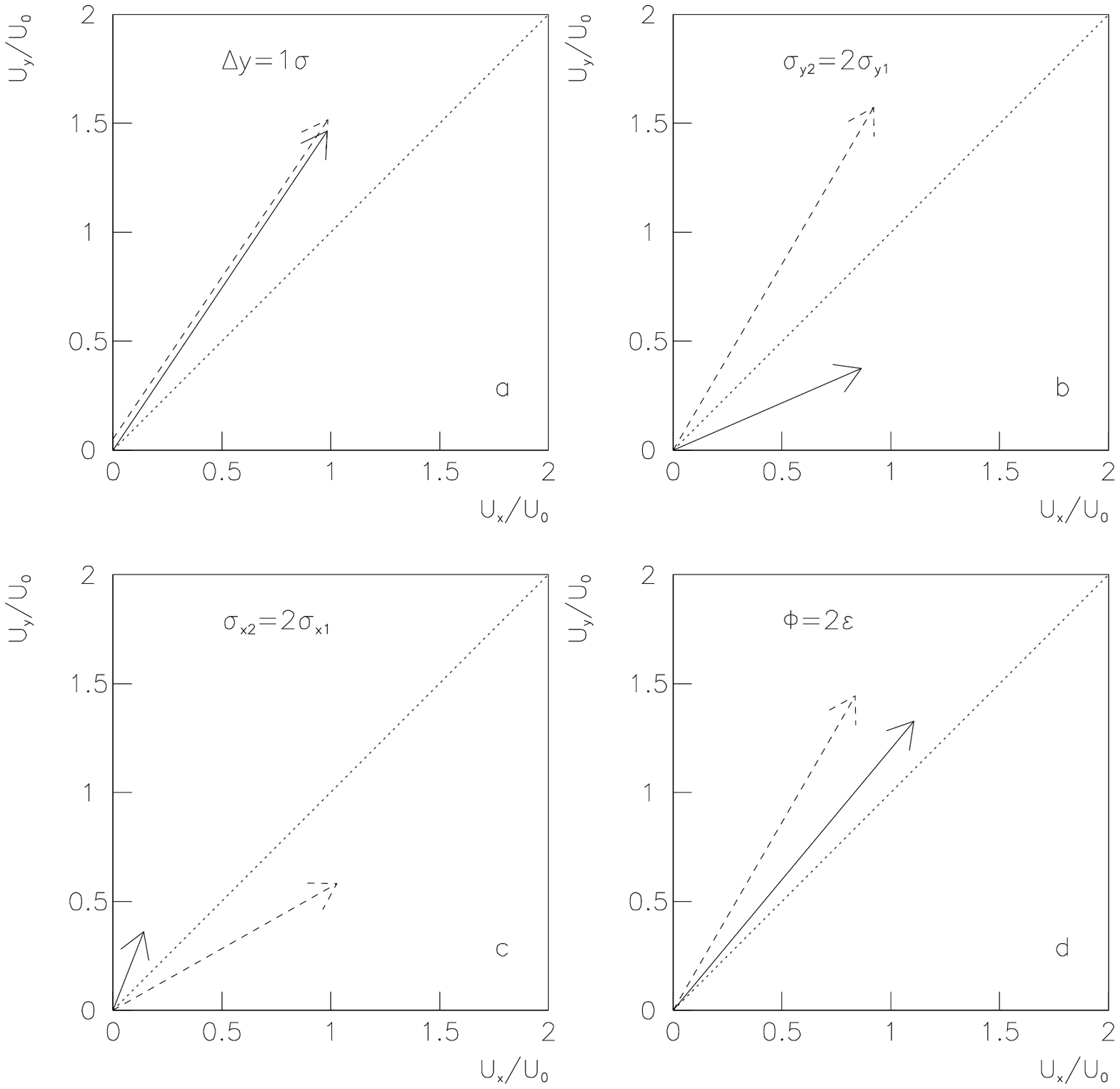,height=5.0in}
\vspace{0.25in}    
\caption{Beamstrahlung diagrams
corresponding to the four pathologies of Figure~3.
The tips of vectors in part a are displaced for display purposes.
Stiff beams are assumed.}
\end{center}
\label{fig:stiffarrow}
\end{figure}
pathology shown in Figure~3.
Each has a unique pattern, which a feedback
algorithm can discern and correct.  In general, if beam~1 is smaller
in $x(y)$ than beam~2, then it will radiate less energy in $x(y)$.

Figure~6 is the same as Figure~5,
but for dynamic beams. Comparison of the
\begin{figure}[h]
\begin{center}\thicklines
\epsfig{file=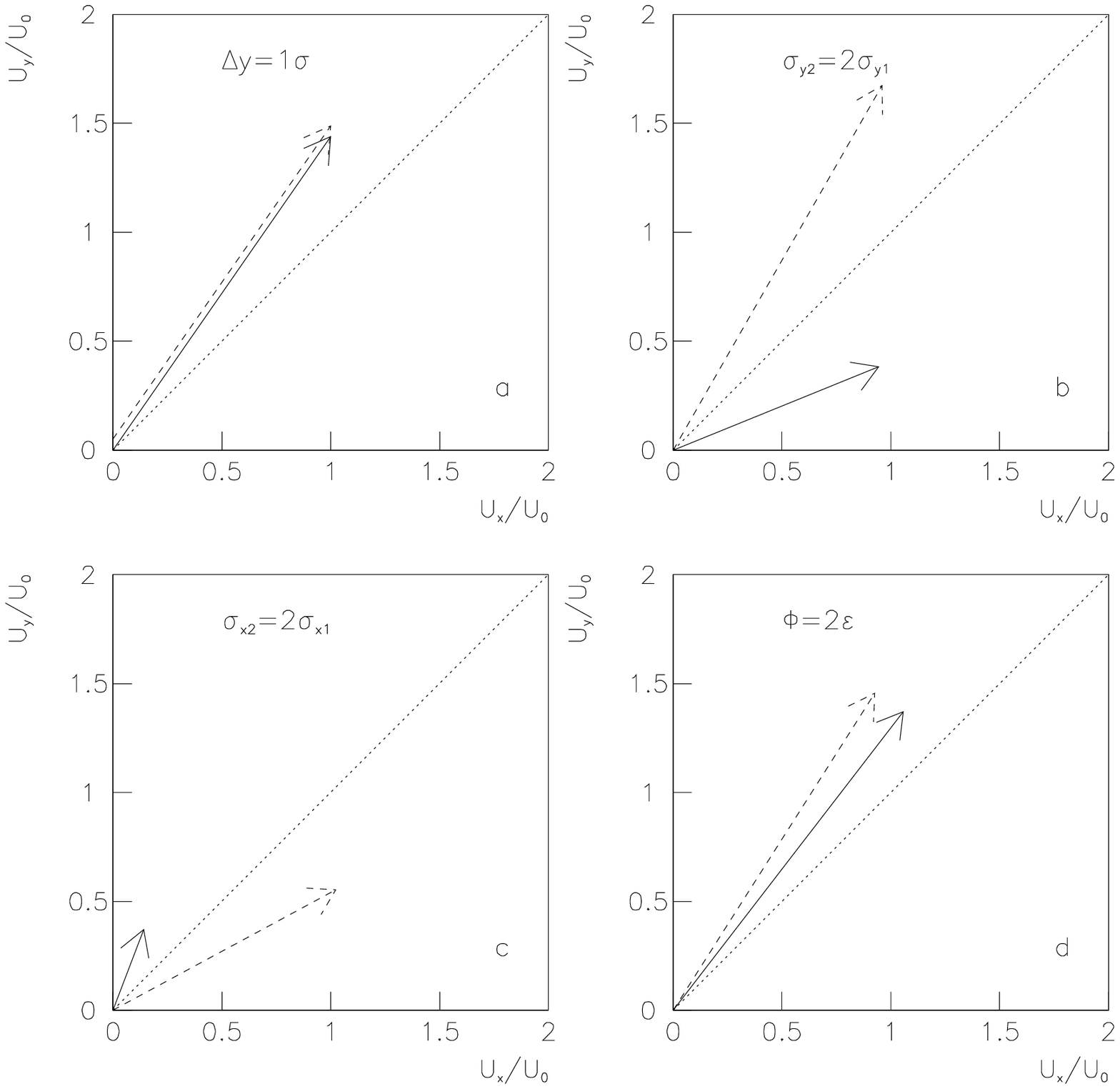,height=5.0in}
\vspace{0.25in}    
\caption{Beamstrahlung diagrams for the same conditions
as Figure 5, but assuming dynamic beams.}
\end{center}
\label{fig:dyarrow}
\end{figure}
two figures shows very little difference.
The effect of dynamic beams is small.
Thus the beamstrahlung diagram presented in this paper is a universal 
display of the pattern associated with beam-beam
pathologies at CESR, PEP-II, KEK, and in the future at a $\sim$1~TeV
$e^+e^-$ machine. 

Figure~7 is the same as
Figures~5 and 6, but with an offset in $x$,
\begin{figure}[h]
\begin{center}\thicklines
\epsfig{file=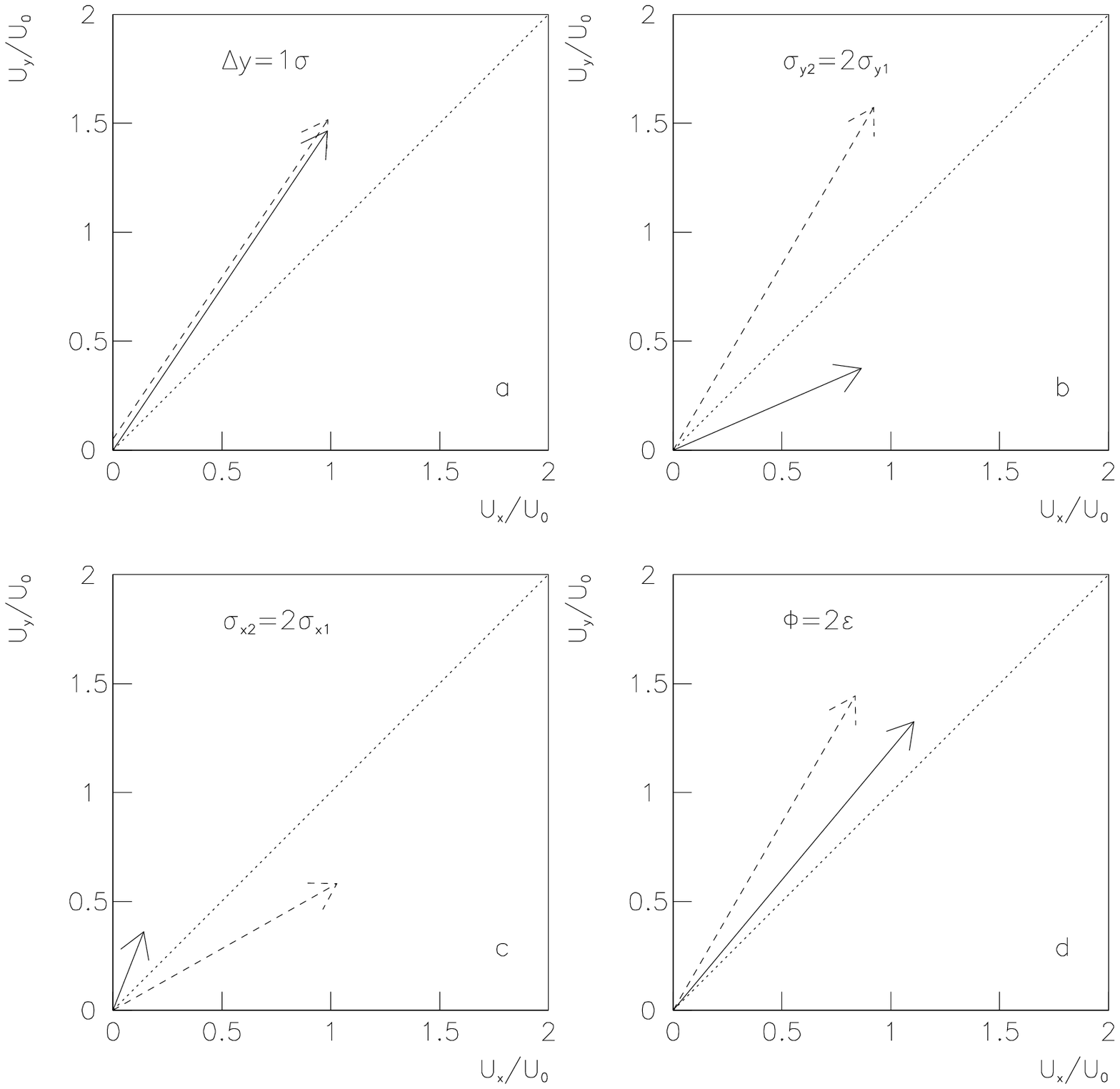,height=5.0in}
\vspace{0.25in}    
\caption{Beamstrahlung diagrams for the same conditions
as Figure 5, but assuming
an $x-$offset of 0.06$\sigma_x$.}
 \end{center}
\label{fig:arrowx}
\end{figure}
18$\mu$m, or 0.06$\sigma_x$, comparable
to the resolution of beam position monitors.
Again very little change is observed with respect to
Figure~5 showing that small horizontal offsets 
have small impact.

Asymmetries corresponding to each pathology in Fig.~3 are defined as
\begin{eqnarray}
A_1 & = & (U_y/U_x-1)\Theta(U_y/U_x-1),\\
\label{eq:a1}
A_2 & = & (U_{2y}/U_{1y}-1)\Theta(U_{2y}/U_{1y}-1),\\
\label{eq:a2}
A_2'& = & (U_{2x}/U_{1x}-1)\Theta(U_{2x}/U_{1x}-1),\\
\label{eq:a2'}
A_3 & = & |\sin{({\bf U}_1,{\bf U}_2)}|,
\label{eq:a3}
\end{eqnarray}
where $\Theta$ is the Heaviside function meaning in this case that the
asymmetries $A_i$ are not defined when the argument of the Heaviside
function becomes negative.
The indexing was chosen to indicate
that the second, a beam bloated vertically, and third, a beam bloated
horizontally,
pathologies are generated from both a zero dipole moment and a
non-zero quadrupole moment in the
transverse charge distribution, and as such
they should be equally ranked.

These asymmetries are not independent.
The usefulness of these beamstrahlung asymmetries is shown in
Figure~8  which displays their dependence
\begin{figure}[h]
\begin{center}\thicklines
\epsfig{file=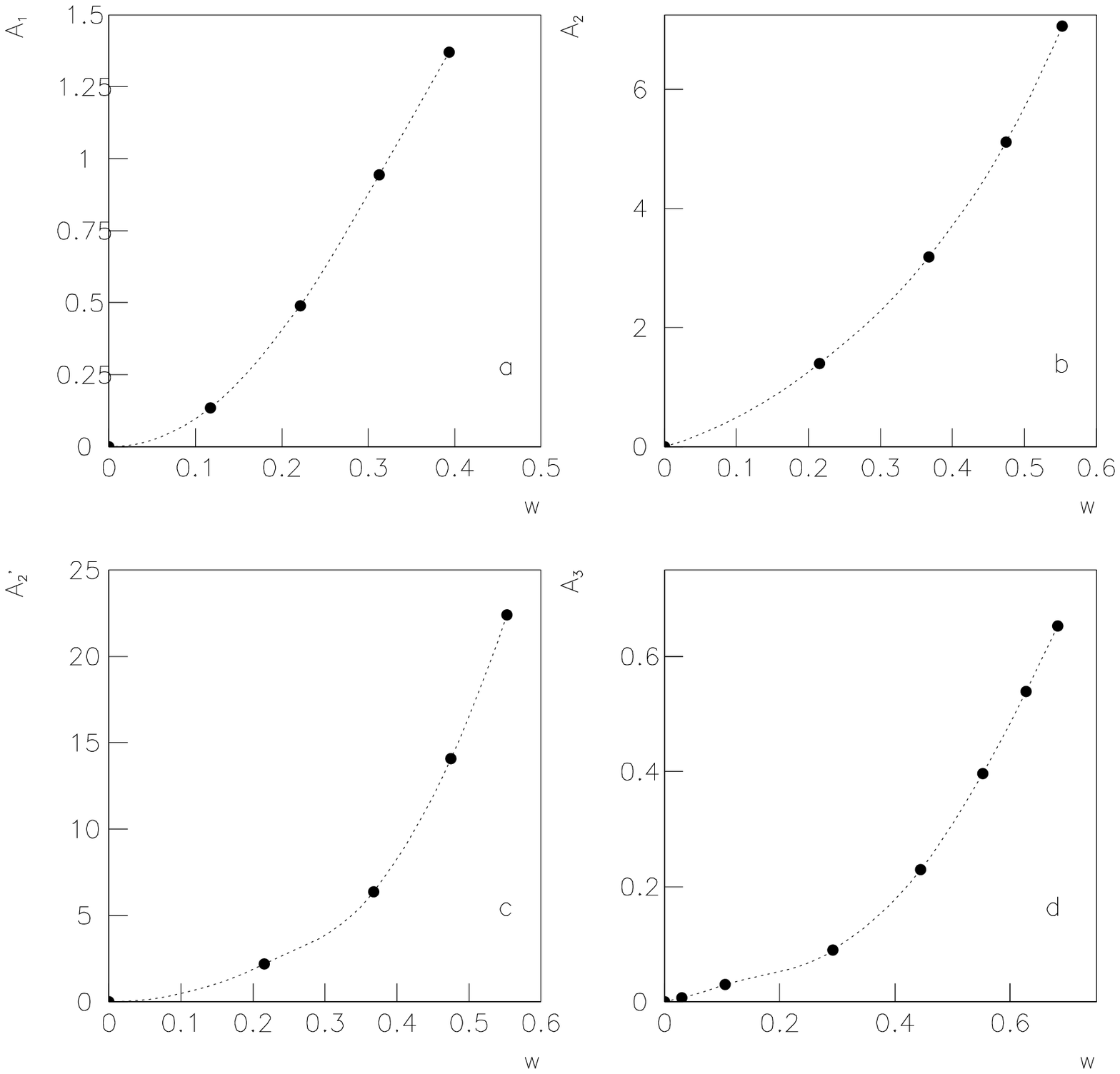,height=5.0in}
\vspace{0.25in}
\caption{Functional dependence of the beamstrahlung asymmetries defined
in the text versus the waste parameter of Equation~3.}
 \end{center}
\label{fig:asymm}
\end{figure}
on the waste parameter defined in Equation~\ref{eq:waste1}.
Each asymmetry's contribution
to the waste parameter of Section~1 is
\begin{equation}
\label{eq:waste}
w_i \sim {\partial w\over \partial A_i} A_i.
\label{eq:awas}
\end{equation}
but they can not be summed together because they are not independent.

Evidence is provided in the next Section that the asymmetries should be 
minimized strictly in the order defined by Eqs.~12-15.
In short, the total waste parameter can be defined as
\begin{equation}
w\sim \sum_i \hat{\partial} w_i A_i,
\end{equation}
where the hatted derivative is defined as
\begin{equation}
\hat{\partial} w_i =({\partial w\over 
\partial A_i})_{A_j=min., j<i}.
\end{equation}
Eq.~17 represents the main result of this paper. The derivatives are
computed, the asymmetries are measured, and the waste parameter is obtained.
Note that if the asymmetries were completely
independent, the specifications $A_j=min.$ would have been unneeded.
Asymmetries 2 and 2' represent both quadrupole corrections, and can be 
interchanged without harm.

For horizontal offsets between the two beams an asymmetry
\begin{equation}
A_1'=(U_x/U_y-1)\Theta(U_x/U_y-1)
\label{eq:a1'}
\end{equation}
can be defined.

We note that for a 10\% change in 
luminosity, the values of the asymmetries change by 
$0.1$ for $A_1$ and $A_1'$, $0.25$ for $A_2$ and $A_2'$ and 
$0.05$ for $A_3$.  Thus these asymmetries
have excellent sensitivity to wasted luminosity.

\section{The Virtual Operator}

Here examples are shown of how the beamstrahlung diagram and
the asymmetries defined in Equations~12-15 and~\ref{eq:a1'}
can by used to eliminate wasted luminosity even in
the presence of multiple pathologies in the beam-beam collision.

We demonstrates this by studying the complete set of
six double pathologies, shown in Figure~9,
which can be derived from the four
\begin{figure}[h]
\begin{center}\thicklines
\epsfig{file=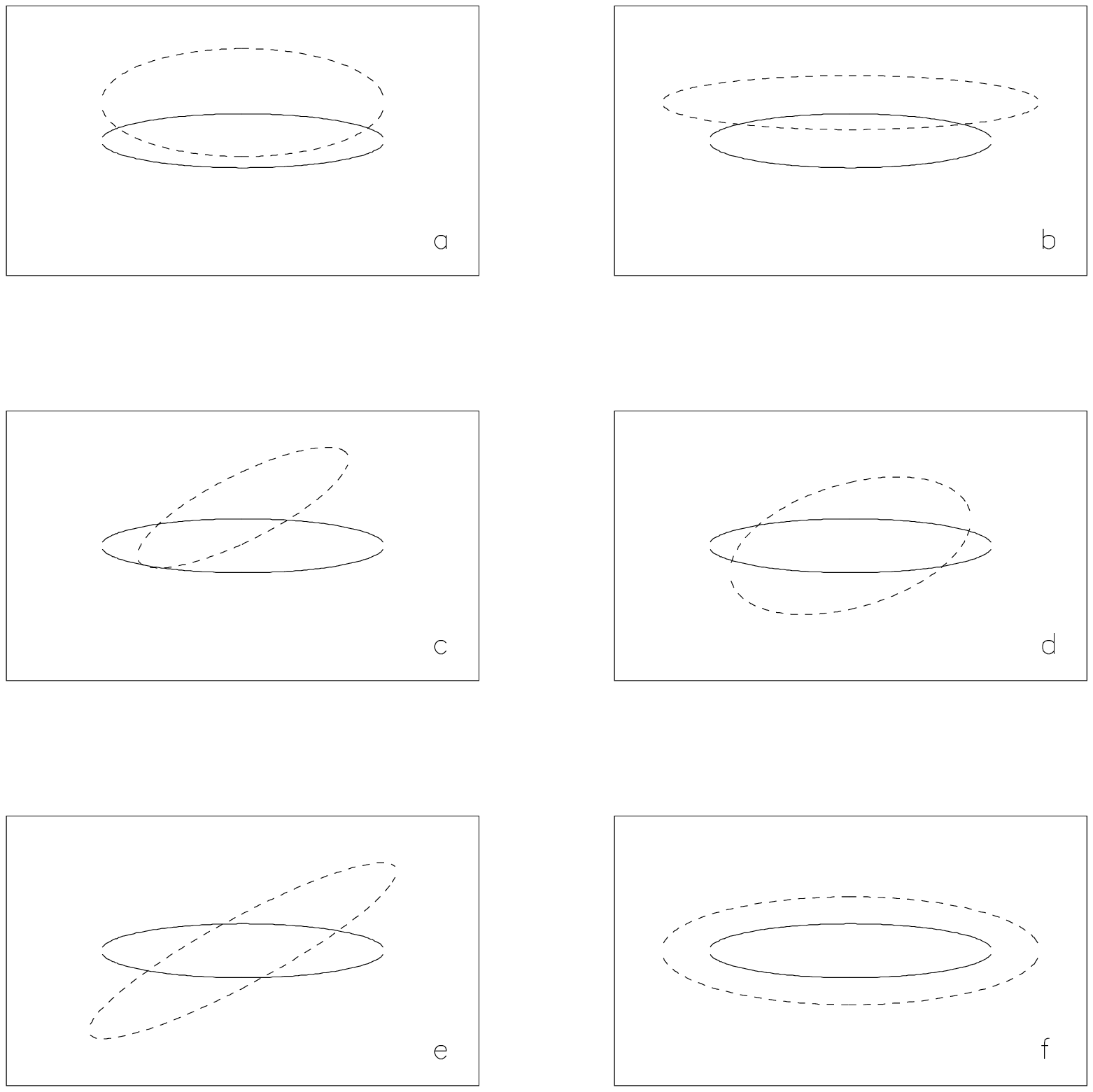,height=5.0in}
\caption{The six possible configurations arising from combinations of any
two of the pathologies of Figure 3. a) $y-$offset and $y-$bloating. 
b) $y-$offset and $x-$bloating.
c) $y-$offset and beam-beam rotation.
d) $y-$bloating and beam-beam rotation.
e) $x-$bloating and beam-beam rotation.
f) $y-$bloating and $x-$bloating.}
 \end{center}
\label{fig:6path}
\end{figure}
single pathologies shown in Figure~3. 
Figure~10 represents the beamstrahlung diagrams corresponding
\begin{figure}[h]
\begin{center}\thicklines
\epsfig{file=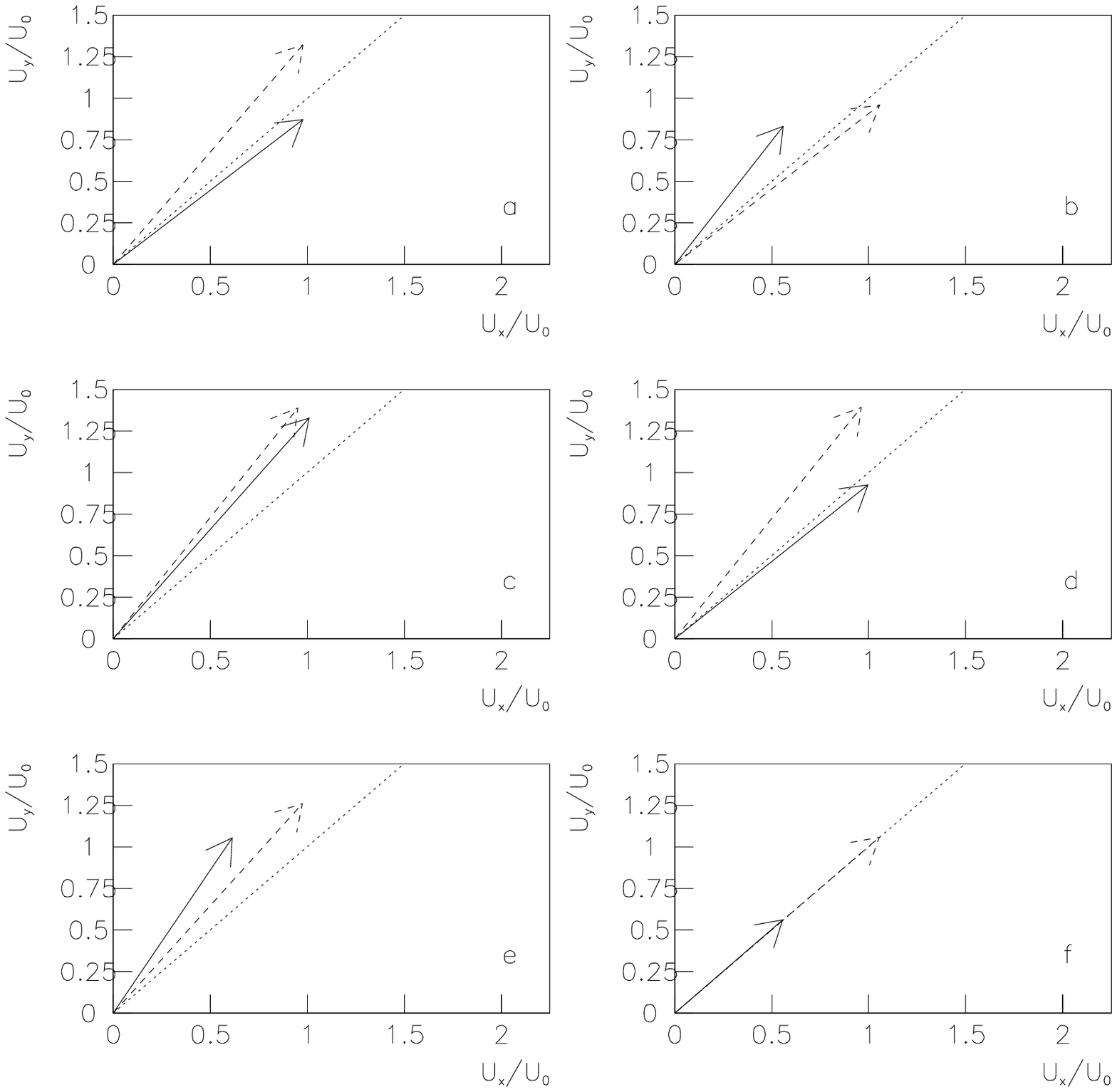,height=5.0in}
\vspace{0.25in}
\caption{Beamstrahlung diagrams corresponding to Figure~9.}
 \end{center}
\label{fig:6arrow}
\end{figure}
to the pathologies
displayed in Figure~9.  
A feedback program,
dubbed the Virtual Operator, finds the highest-ranking 
asymmetry, minimizes it
by changing the appropriate collision parameter, and
obtains the beamstrahlung diagrams of Figure~11,
displaying only one pathology
\begin{figure}[h]
\begin{center}\thicklines
\epsfig{file=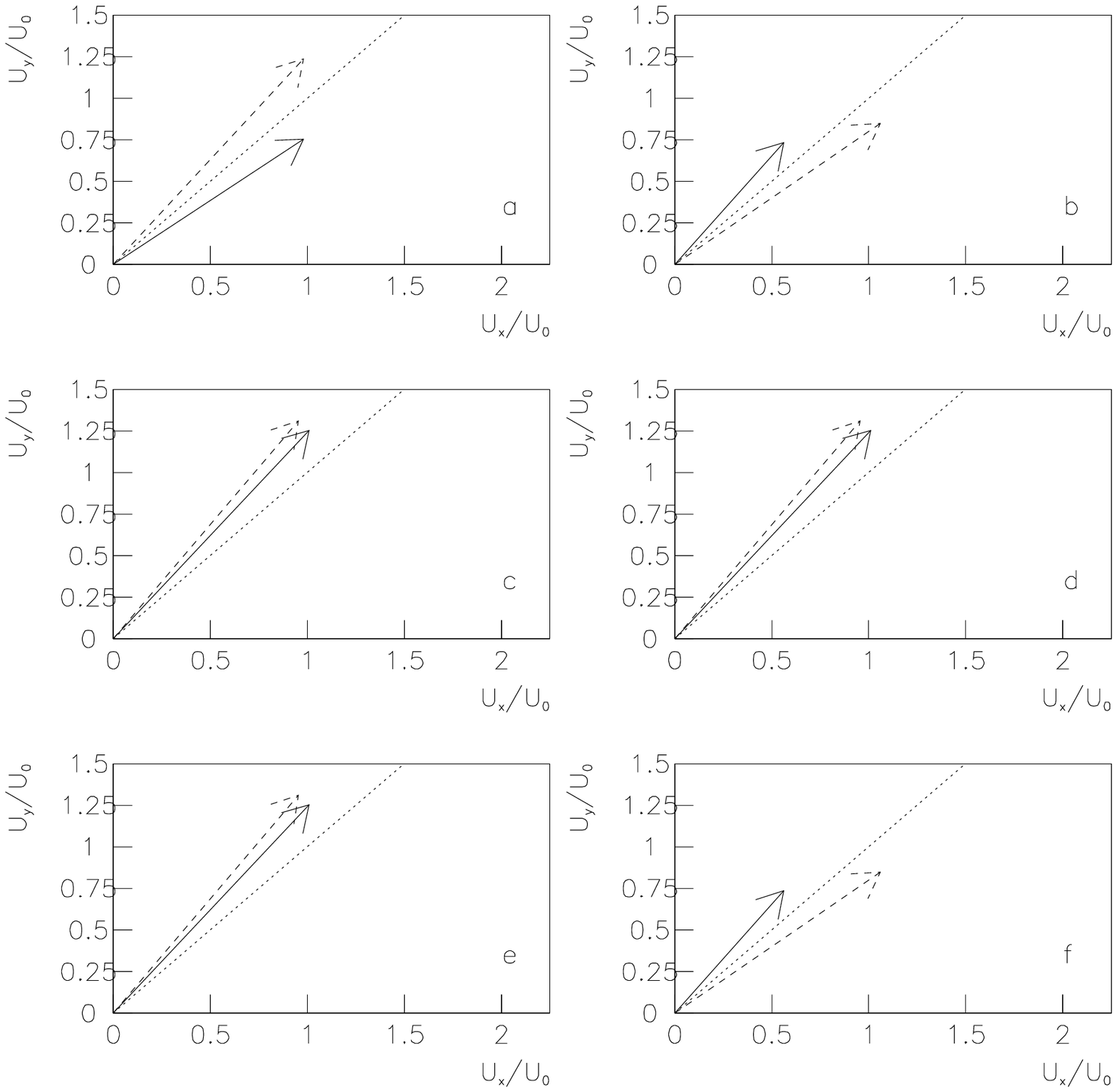,height=5.0in}
\vspace{0.25in}
\caption{Beamstrahlung diagrams, corresponding to Figures~9
and 10, after
correction of the dominant asymmetry. Compare with Figure~3.}
\end{center}
\label{fig:6corr}
\end{figure}
which is trivial to correct. 

Two comments are in order. First, if the largest asymmetry were to be
corrected first, instead of the highest-ranked, convergence would not 
be achieved. To prove the point, it is enough to compare Figs.~5b, 10d and
11d. If the sextupole correction is acted upon first, $A_3$ would have to
increase as opposed to being minimized.

Second, we wish to prove that minimization of a higher-ranked
asymmetry effectively corrects the associated pathology. Although all the
double pathologies were tried, only Fig.~9c,
which corresponds to a vertical offset plus a rotation, is presented.
$A_1$ and $A_3$ are the two most correlated asymmetries. 
The asymmetry $A_1$ is not zeroed,
and cannot be zeroed by moving one beam.
Figure~12 shows the dependence of
$A_1$ versus the vertical offset, showing
\begin{figure}[h]
 \begin{center}\thicklines
\epsfig{file=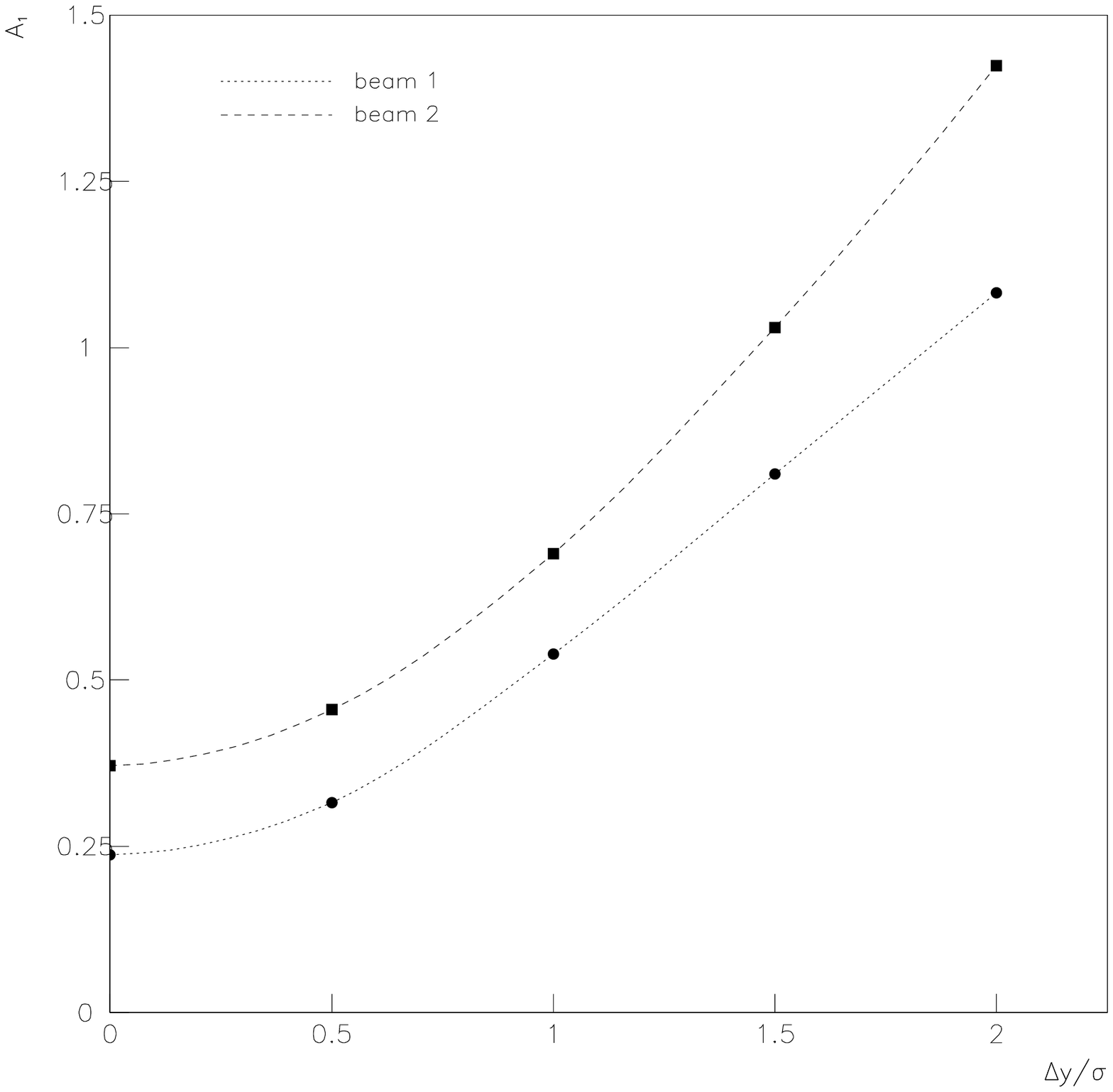,height=5.0in}
\vspace{0.25in}
\caption{The dependence of the first asymmetry $A_1$, as defined in the
text, versus the vertical offset for the case of a vertical offset
plus a rotation.}     
 \end{center}
\label{fig:a1off}
\end{figure}
that minimization of the asymetry gives the desired correction.

We did not consider horizontal offsets here, but they could
easily have been included, as discussed in the previous Section.

\section{Conclusion}

The beamstrahlung diagram and asymmetries derived here
demonstrate a complete and rigorous method for luminosity 
optimization. The wasted luminosity is for the first time
related to quantities that are instantaneously observable,
and specify the necessary correction.
We have considered a complete class of beam-beam pathologies.

If the machine is perfectly symmetric a beamstrahlung monitor is
very useful for measuring the size of the beam.
In the case of asymmetric beams a beamstrahlung monitor is extremely
powerful.  The study of the beamstrahlung diagram derived from
the power and polarization of the beamstrahlung signal
allows identification of the beam-beam pathology, identification of the 
``bad'' beam, and
measures the correction that needs to be applied.
In short the wide angle beamstrahlung signal analyzed in the manner
described here is a powerful tool to eliminate wasted luminosity
at present and future $e^+e^-$ colliders.

\section*{Appendix A}

The properties of short magnet radiation were first discussed by Coisson
in reference \cite{coisson}. In the classical model\cite{jackson}, 
the bent electron is made to
sweep through the detector in a ``searchlight'' fashion,
effectively covering all beam-detector angles. In the Coisson's model
the opposite extreme is adopted, and
the angle is kept constant throughout the orbit, the large angle approximation.
Both models predict the same power, the same total polarization,
and the same typical angle, of order $1/\gamma$ for the emitted radiation,
but they differ dramatically in the spectrum at large angles.

The Coisson model is of interest here because the detector's angle is
constant throughout the collision at colliders such as CESR.
At large angle the classical model predicts a steep fall-off of the
power, exponential both in the photon energy and in the cube of
the observation angle.
The Coisson model predicts 
three properties of large angle beamstrahlung radiation.
They are:
\begin{itemize}
\item The cutoff energy, at large angle, does not depend on $\gamma$.
There is no exponential fall-off as predicted by the
``searchlight'' approximation, making
detection possible. In particular at 6 mrad at CESR, for example,
visible radiation is at or below the cutoff frequency.
\item The polarization is linear at a fixed location
in azimuth with an eightfold pattern, $(\cos^2{2\phi},\sin^2{2\phi})$
around the azimuth. The angle $\phi$ is the angle between the net transverse
force experienced by the beam and the detector location. Thus
the pattern of the polarization provides information
about the beam-beam overlap.
\item The large angle double differential spectrum is proportional to 
$(\gamma\theta)^{-4}$, and not exponential.  The large angle
power scales as $1/\gamma^2$. Thus the situation at 
$B$ factories is more favorable than at higher energy machines.
\end{itemize}

These properties are re-derived here in an elementary way for
constant large angle of detection.
Consider an extremely relativistic particle, $\gamma>>1$, 
undergoing a vertical
deflection, due for example to a horizontal dipole magnet
exerting a force $F$ over a length $\sigma_z$. 
Radiation of energy $k=h\omega$ is  
detected at an angle $\theta$ which is much larger than $1/\gamma$.
In the laboratory frame the radiated energy
is equal to \cite{jackson}
\begin{equation}
U = {2\over 3}{r_e\over mc^2}\gamma^2F^2\sigma_z.
\label{eq:jacku}
\end{equation}
A simpler derivation is possible by studying the radiation in the
rest frame of the radiating particle.  Note that
all quantities in the particle rest frame are starred as
shown in Figure~13.
\begin{figure}[h]
\begin{center}\thicklines
\epsfig{file=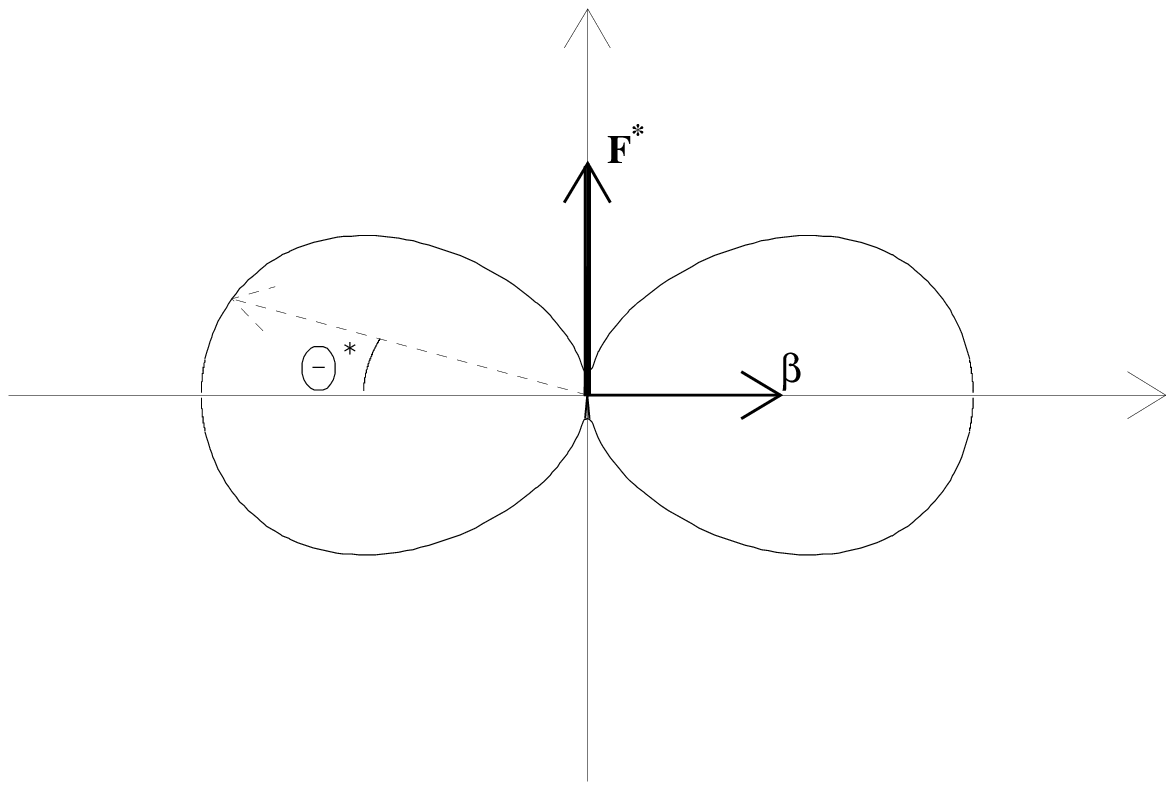,height=5.0in}
\caption[]{Dipole radiation in a radiating particle's rest frame.
Indicated are
the direction of the force and the angle corresponding to the observation
angle in the laboratory frame. }
\end{center}
\label{fig:raddef}
\end{figure}
The radiation will have a dipole pattern with angular intensity
proportional to the squared sine of the angle between the direction of
detection and the direction of the force. The force
maintains its vertical direction and has a modulus 
\begin{equation}
F^*=2\gamma F.
\label{eq:F*}
\end{equation}
The angle is very large in
the laboratory frame, and the corresponding direction in the rest frame is
very close to the backward direction. In a perturbative treatment
the angle $\theta^*$ is taken with respect to the direction opposite the
direction of motion (Fig.~13).
If only small angular components along the direction 
of the force are considered 
\begin{equation}
I(\theta^*)\propto \cos^2{\theta^*}.
\label{eq:int}
\end{equation}
The intensity is essentially constant at small angles in the rest frame.

The relation between the energies and angles in the lab
and radiating particle rest frames is given by
\begin{eqnarray}
k      & = & {k^*\gamma\theta^{*2}\over 2},\\
\label{eq:kdef}
\omega & = & {\omega^*\gamma\theta^{*2}\over 2},\\
\label{eq:omegadef}
\theta & = & {2\over\gamma\theta^*}.
\label{eq:thetadef}
\end{eqnarray}
The direction of the radiation in the
radiating particle rest frame is at CESR, for example,
$\theta^*\sim 0.03$ or two degrees away from the backward axis.

In the radiating particle rest frame 
the cutoff frequency is inversely proportional to the duration of the
perturbation, which is $\sigma^*_z/c$. Using equations 24 and 25,
and the relativistic formula length-dilation, the following relations
are obtained
\begin{eqnarray}
\sigma^*_z &=& \sigma_z/\gamma,\\
\label{eq:sigz}
\omega^*_c&\sim& O({c\over\sigma^*_z})\\
\label{eq:omega*}
\omega_c &\sim& O({c\over \sigma_z\theta^2}),
\label{eq:omegac}
\end{eqnarray}
which shows that
the cutoff frequency at large angle does not depend on
$\gamma$, the first prediction by the Coisson model. 
At CESR, $\omega_c \sim 10^{16}$sec$^{-1}$, which 
is of order of the visible light frequency.

The polarization vector of the emitted
radiation in the radiating particle rest frame
is given by \cite{jackson}
\begin{equation}
{\bf E^*}(R)={e\over m Rc^2}{\bf n^*}\times ({\bf n^*}\times {\bf F^*}),
\label{eq:E*}
\end{equation}
where ${\bf n}^*$ is the unit vector along the direction of observation.

Using Equation~\ref{eq:E*}, 
and the condition of orthogonality between ${\bf E^*}$, 
${\bf B^*}$ and ${\bf n^*}$, the three vectors are 
\begin{eqnarray}
{\bf E^*} & = & K(\theta^{*2}\sin{\phi}\cos{\phi},
                  \theta^{*2}\sin^2{\phi}-1, -\theta^* \sin{\phi})\\
\label{eq:E*vec}
{\bf B^*} &=& K(-1+\theta^{*2}/2, 0, -\theta^* \cos{\phi}),\\
\label{eq:B*vec}
{\bf n^*} &=& (\theta^* \cos{\phi}, \theta^* \sin{\phi}, -1+\theta^{*2}/2),
\label{eq:n*vec}
\end{eqnarray}
with $K$ a constant.
The polarization component along $x$ and $y$ in the laboratory frame are
\begin{eqnarray}
 E_x & = & \gamma(E^*_x-B^*_y)=+K{\gamma\theta^{*2}\over 2}
           \sin{2\phi}=+K{2\over \gamma\theta^2}\sin{2\phi}\\
\label{eq:Ex}
 E_y & = & \gamma(E^*_y+B^*_x)=-K{\gamma\theta^{*2}\over 2}
           \cos{2\phi}=-K{2\over \gamma\theta^2}\cos{2\phi}.
\label{eq:Ey}
\end{eqnarray}
Thus each component has four azimuthal zeros, and information is replicated
every 45 degrees, which is the second prediction of the Coisson model.

The total energy radiated in the
laboratory frame can be expressed as an average over the boosted photon 
energies in the rest frame, times the number of photons $N$
\begin{equation}
 U= \sum \gamma(k^*+k^*_z) = N 
<\gamma(k^*+k^*_z)>=N\gamma<k^*>.
\label{eq:Ugam}
\end{equation}
The energy flowing into a detector covering a solid angle $d\Omega$, 
located at large 
angle, can be easily computed in the radiating particle frame. Using 
$I(\theta^*)\sim 1$, Equations 23 and 25,
and neglecting factors
of order one an expression for the large angle spectrum is obtained,
\begin{equation}
 \Delta U \sim N\theta^*
\Delta\theta^*\Delta\phi {\gamma <k^*>\theta^{*2}\over 2}
 ={8U\over\gamma^4\theta^5}\Delta\theta\Delta\phi.
\label{eq:delU}
\end{equation}
The energy in the lab frame, $U$, contains a dependence on $\gamma^2$.
The angular factor integrates to a constant 
(which agrees with Equation~8 in Ref.~\cite{welch}), 
leaving the $1/\gamma^2$ dependence.
This is purely due to kinematics.
At CESR, for example,
10nW of visible beamstrahlung are available between 6 and 7 mrad.

\section*{Appendix B}

A beam-beam
interaction simulation was developed from the program
described in reference \cite{holleb}. 
Gaussian beams in all three dimensions
are assumed.
Beams are sliced in 3-dimensional cells.
The cells are typically 0.25-0.5$\sigma$ along each axis
and extend out to 3-4$\sigma$ in each direction.
Thus a total of $10^3$ to $3\times 10^4$ cells are simulated. 
The beams are then made to cross each other. In the first
step, the first layer of the positron beam encounters the 
first layer of the electron beam. The electric fields are purely 
transverse to $O(1/\gamma)$, and are computed assuming that the charge is 
located in a sphere located in the center of the cell.
This is the ``cloud-in-cell'' model.
Assuming cylindrical coordinates, 
a cell in beam one gets a total transverse 
deflection \cite{jackson1}
\begin{equation} 
\Delta{\bf r'}_{1j}= -{2N_2r_e\over\gamma}\sum {P_{2i}{\bf b}_{ij}
\over b_{ij}^2}.
\label{eq:deltar'1}
\end{equation}
The summation runs over the cells in the opposite layer, ${\bf b}_{ij}$ is
the impact parameter between cell $j$ in beam one and cell $i$ in beam two,
and $P_{2i}$ is the fraction of charge in cell $i$.
At the end of each layer-layer interaction positions and velocities are 
updated, 
\begin{eqnarray} 
{\bf r'}_{j}&=& {\bf r'}_{j}+\Delta{\bf r'}_{j},\\
\label{eq:r'}
{\bf r}_{j}&=& {\bf r}_{j}+{\bf r'}_{j}\Delta z.
\label{eq:r}
\end{eqnarray}
$\Delta z$ is the unit step taken along the beam direction.
This allows for dynamic beams, with each beam pinching
the other as the collision progresses, and the luminosity is computed
as an overlap of the dynamic density functions.

The program of 
reference \cite{holleb} was found to be unfit for the simulation of 
flat beams.
If the lattice is chosen to have the same number of cells in each dimension, 
the cells will be as flat as the beam. If the charge is then
concentrated in the centers, a large force will be calculated, where in reality
the total force is small, due to the cancellations of the large $x$ components
in the integral over the cells.  Figure~14 illustrates this.
\begin{figure}[h]
\begin{center}\thicklines
\epsfig{file=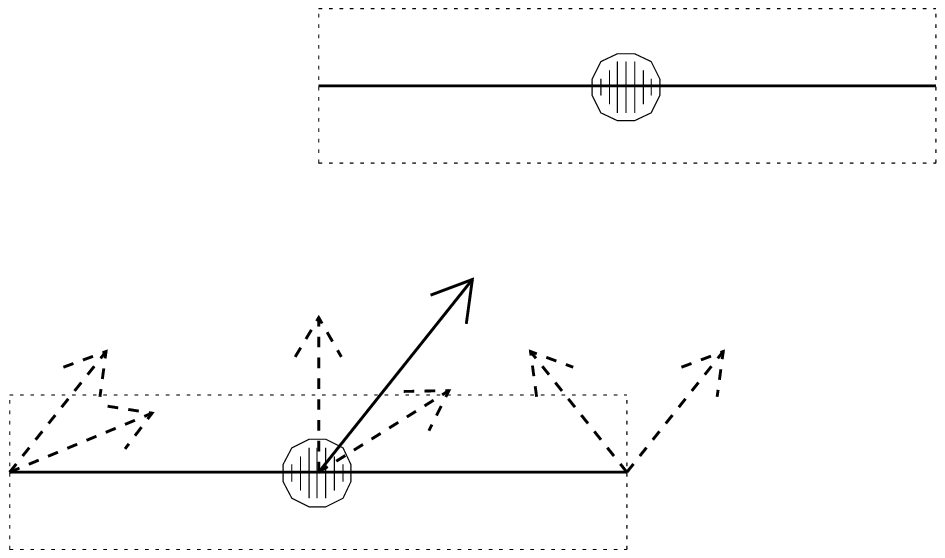,height=5.0in}
\caption[]{Cell-cell interaction in the simulation program. The cell has
an aspect ratio similar to the beam aspect ratio. In the ``cloud-in-cell'' 
model, all the charge is concentrated in a point in the center of the cell.
In the ``matchstick-in-cell'' model, the charge is spread over a line along the
cell.}
\end{center}
\label{fig:match}
\end{figure} 

To reduce this problem the number of cells in $x$ should be enlarged 
to make each cell square in the transverse plane. This solution is very 
CPU time-consuming.  A solution was found by replacing each cell with a line
of charge, called a ``matchstick,'' and computing the integral
\begin{equation}
 \Delta{\bf r'}_{ij}= {-2N_2r_e P_i\over\gamma} 
\int dx_i dx_j {{\bf b}_{ij}\over  b_{ij}^2}.
\label{eq:deltar'2}
\end{equation}
For the purpose of improving the convergence of the program, the 
matchsticks were kept horizontal throughout the interaction. Assuming
matchstick lengths $L_i$ and $L_j$, the solution to the integral above is
\begin{equation}
 \Delta{\bf r'}_{ij}= 
{-2N_2r_e P_i\over\gamma} (\sum_1^4 t_n f_n +2b_yg_n,
\sum_1^4 2t_n g_n -b_yf_n),
\label{eq:deltar'3}
\end{equation}
where
\begin{eqnarray}
t_1 & = & b_x+{L_i+L_j\over 2} \\
t_2 & = & b_x+{L_i-L_j\over 2} \\
t_3 & = & b_x-{L_j+L_i\over 2} \\
t_4 & = & b_x+{L_j-L_i\over 2} \\
f_n & = & (-1)^{n+1}\log{(t_n^2+b_y^2)}\\
g_n & = & (-1)^{n+1}\tan^{-1}{(t_n/b_y)}.
\end{eqnarray}
Given the deflection vector, ${\bf r'}$, the total radiated energy
with both $x-$ and $y-$ polarization is computed 
using
\begin{equation} 
{\bf F}={\gamma m c^2\over 2\Delta z}\Delta{\bf r'}.
\end{equation}
The energy vector ${\bf U}$ for each beam is computed by summing
\begin{eqnarray}
U_x = \sum\Delta U_{xj}&=&\sum{2N\over 3mc^2}P_j r_e\gamma^2F_x^2\Delta z, \\
U_y = \sum\Delta U_{yj}&=&\sum{2N\over 3mc^2}P_j r_e\gamma^2F_y^2\Delta z.
\end{eqnarray}

The program continues to interact the beams, layer by layer, 
updating trajectories with Equations~38-39, until
the beams fully cross each other.
An option was inserted in the program to use or not to use Equations~38-39,
that is to make the beams either dynamic or stiff.
The reason for the option was to compare against
existing analytic predictions for beamstrahlung given
in reference \cite{albert}.

\section*{Appendix C}

The simulation program described above is
used to make comparisons with
the analytic predictions of reference~\cite{albert} which
are valid only for stiff beams. A slow, quadratic convergence was found when
diminishing the cell size. When a cell of one beam overlaps with one from
the other beam, the program computes a zero field. In reality, the contribution
of nearby particles is important, due to the $1/b$ dependence of the field.
Because the emitted power depends on the field squared, the dominant
convergence is quadratic.

To adjust for this fact using finite computer resources
two different lattice sizes $a$ and $b$ were used.
The ``exact''
emitted energy $U_{ex}$ was extracted
using the linear system
\begin{eqnarray}
U_a & = & U_{ex} -\alpha a^2,\\
\label{eq:ua}
U_b & = & U_{ex} -\alpha b^2,
\label{eq:ub}
\end{eqnarray}
and solved for $U_{ex}$ and $\alpha$.

\begin{table}
\begin{tabular}{|c|c|c|c|c|c|}\hline
Quantity& Bin$=0.25\sigma$ & Bin$=0.3\sigma$ & Fit  & Analytic
& Dyn.beams\\
 \hline
$U_{x}(10^{12}$eV) & .4002 & .3979  
&.4055 & .4051 & .4088\\
$U_{y}(10^{12}$eV) & .4013  & .3997  
& .4049 & .4051 & .4163\\
$L/L_0$   & 1.00  & 1.00 & 1.00 & 1.00 & 1.12\\
\hline
\end{tabular}.
\caption{Comparison between two different binnings, the fitted values, 
according to Equations~51-52,
and the analytic predictions, all for stiff
beams. The last column shows the same quantities for dynamic beams. }
\label{tab:comp}
\end{table}
The stiff-beam comparison of ${\bf U}$, with different cell sizes and  
against the analytic predictions of reference~\cite{albert} using
the beam parameters of Table~\ref{tab:para} are shown in 
Table~\ref{tab:comp}. There is agreement between analytic and
simulation at the 0.2\% level.
If the beams are dynamic particles
will move during the
collision by about 1~micron, leaving the beam virtually unchanged in $x$ but
generating a substantial ($O(10\%)$) squeezing in $y$. The squeezing will 
have two effects: it will increase the luminosity
and it will generate slightly
more power. There will also be a slight asymmetry between $x$ and $y$ and some
net polarization.
The luminosity increases by 12\%. The luminosity calculation was checked, for 
round beams, against the program of reference~\cite{holleb} and our
simulation agrees to within 1\%.
Figure~15 shows the 
\begin{figure}[h]
\begin{center}\thicklines
\epsfig{file=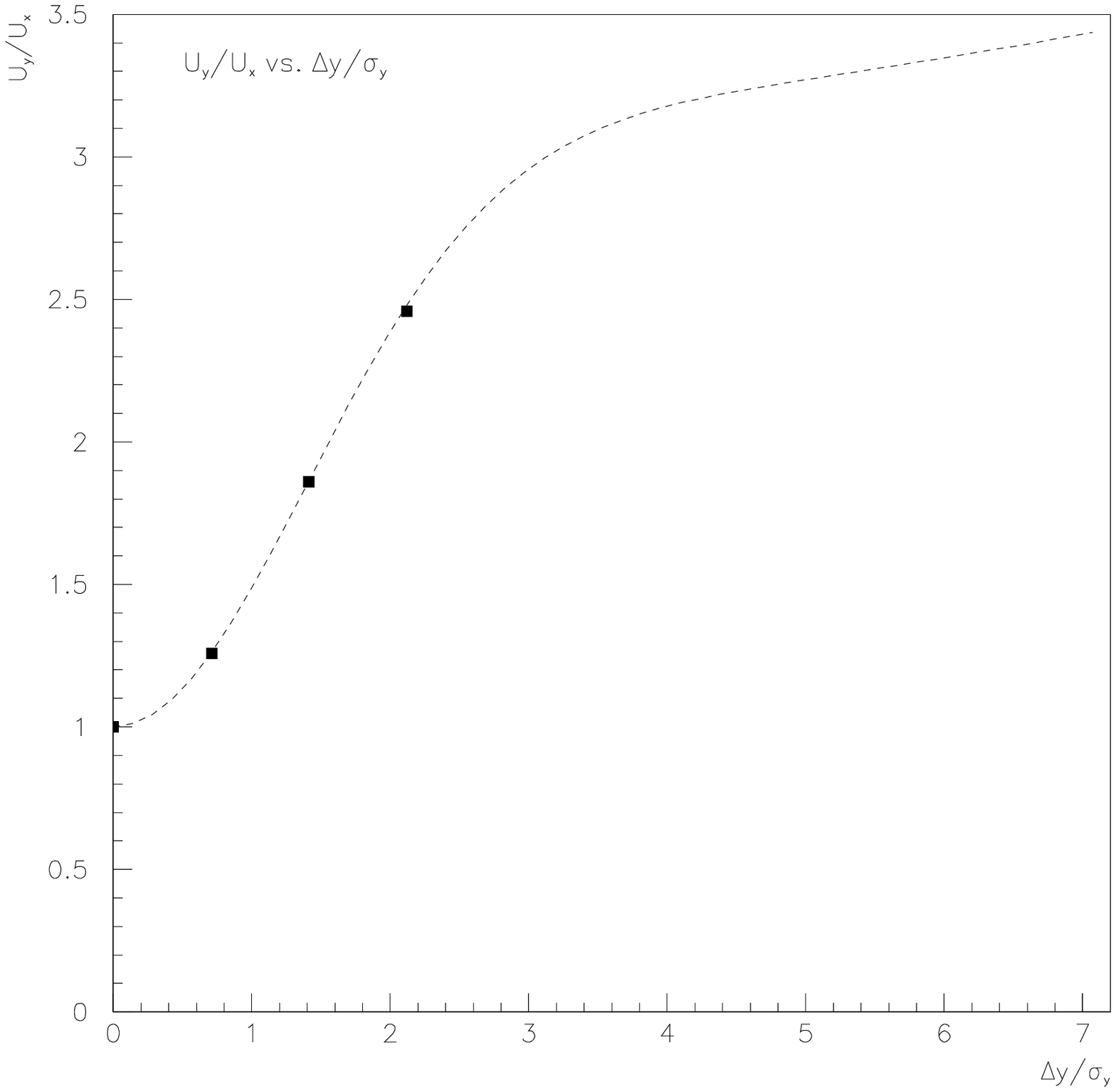,height=5.0in}
\vspace{0.25in}
\caption[]{Radiation polarization versus beam-beam offset. The solid line
is the analytic prediction from reference~\cite{albert}, and the dots
are from the simulation described in the text.}
\end{center}
\label{fig:comp}
\end{figure}
analytical versus simulation comparison of $U_y/U_x$ when two flat
beams are separated by a vertical offset.
We conclude that our simulation
method has a precision of order few per thousand for beamstrahlung 
computations.

\end{document}